\documentclass[aps, showpacs,preprintnumbers,amsmath,amssymb]{revtex4-1}

\usepackage{amsmath,amssymb,bm}
\usepackage[dvips]{graphicx}
\usepackage{yfonts}
\usepackage{enumerate}
\usepackage{bm}
\usepackage{color}

\newcommand{\tr}{\mathop{\mathrm{tr}}}
\newcommand{\bl}{\biggl}
\newcommand{\br}{\biggr}
\newcommand{\vvL}{\delta \hat{\bm{v}}_L}
\newcommand{\vvE}{\delta\hat{\bm{v}}_E}
\newcommand{\vL}{\delta \hat{v}_L}
\newcommand{\vE}{\delta \hat{v}_E}
\newcommand{\deln}{\delta \hat{n}}
\newcommand{\dele}{\delta \hat{e}}
\newcommand{\delp}{\delta \hat{p}}
\newcommand{\delj}{\delta \hat{j}}
\newcommand{\vdelp}{\delta \hat{\bm{p}} }
\newcommand{\vdelj}{\delta \hat{\bm{j}} }
\newcommand{\cP}{{\cal P}}
\newcommand{\cQ}{{\cal Q}}
\newcommand{\average}[1]{\langle#1\rangle_\text{eq}}
\newcommand{\noise}{\hat{R}}
\newcommand{\memory}{\varPhi}
\newcommand{\Lv}{\hat{\mathcal{L}}}

\newcommand{\hO}{\hat{\mathcal{O}}}
\newcommand{\hH}{\hat{H}}
\newcommand{\hN}{\hat{N}}
\newcommand{\hA}{\hat{A}}
\newcommand{\hB}{\hat{B}}
\newcommand{\hT}{\hat{T}}
\newcommand{\hj}{\hat{j}}
\newcommand{\delbeta}{\delta \hat{\beta}}
\newcommand{\delT}{\delta \hat{T}}
\newcommand{\delP}{\delta \hat{P}}
\newcommand{\delmu}{\delta (\widehat{\beta \mu} )}
\begin{document}

\preprint{RIKEN-QHP-45, RIKEN-MP-55}
\author{Yuki Minami}
\author{Yoshimasa Hidaka}
\affiliation{
Theoretical Research Division, Nishina Center, RIKEN, Wako 351-0198, Japan}

\date{\today}

\title{Relativistic hydrodynamics from projection operator method}
\begin{abstract}
We study relativistic hydrodynamics in the linear regime, based on Mori's projection operator method.
In relativistic hydrodynamics, it is considered that an ambiguity about the fluid velocity occurs
from a choice of a local rest frame: the Landau and Eckart frames.
We find that the difference of the frames is not the choice of the local rest frame,
but rather that of dynamic variables in the linear regime.
We derive hydrodynamic equations in the both frames by the projection operator method.
We show that natural derivation gives the linearized Landau equation.
Also, we find that, even for the Eckart frame, the slow dynamics is actually described by 
the dynamic variables for the Landau frame.
\end{abstract}
\maketitle

\section{Introduction}
\label{sec:intro}
Relativistic hydrodynamics has been widely applied for studying relativistic nonequilibrium phenomena.
For examples, it describes hadron spectra and elliptic flow in the  heavy ion physics~\cite{Hirano,rischke}, and
jets in the astrophysics~\cite{mizuta,maartens}.
The hydrodynamic equations applied to these systems are mainly those for perfect fluids.
One of reasons is that the dissipative effects in relativistic hydrodynamics are not fully understood,
e.g, some pathological problems arise from the dissipative effects:
 the acausal propagation and the instability of the equilibrium state~\cite{hiscock}.
Although many hydrodynamic equations have been proposed to resolve these problems
\cite{israel,Denicol,Van,Muronga,Tsumura:2011cj,Natsuume:2007ty},
it is not still obvious which equation describes the correct behavior of the relativistic dissipative fluid.
Namely,  even the basic equation has not been established in relativistic hydrodynamics.

The relativistic hydrodynamic equations are generally given as the following conservation laws:
\begin{align}
\partial_\mu j^{\mu} &=0, \label{eq:conservation1}\\
\partial_\mu T^{\mu \nu} &=0. \label{eq:conservation2}
\end{align}
Here, $j^\mu$ is the particle current and $T^{\mu \nu}$ is the energy-momentum tensor. They are decomposed  into 
 \begin{align}
j^\mu &= n u^\mu+\nu^\mu, \\
T^{\mu \nu}&=h u^{\mu}u^{\nu}-Pg^{\mu\nu}+q^\mu u^\nu +q^\nu u^\mu +\tau^{\mu\nu},  
\end{align}
where $n$ is the particle density,  $h=e+P$ the enthalpy density, 
$P$ the pressure, $e$ the energy density, $u^\mu$ the fluid-four velocity.
The dissipative terms, $\nu^\mu$, $q^\mu$,  and $\tau^{\mu\nu}$, denote the particle and energy diffusions, and the viscous stress tensor, respectively.
The explicit expressions of these terms are not unique but depending on considered equations. 
This ambiguity comes from a choice of local rest frames of the fluid.

To see this ambiguity, let us classify the hydrodynamic equations into two groups: the Eckart and Landau frames~\cite{landau,eckart}.
In the Eckart frame, the local-rest frame is defined as that of the particle current, i.e., 
the fluid velocity is proportional  to the particle current: 
\begin{equation}
u^\mu_{\rm E} \propto j^\mu .
\end{equation}
In this frame, the particle diffusion is absent, $\nu^{\mu} = 0$.
On the other hand, in the Landau frame, the fluid velocity is proportional to the energy current: 
\begin{equation}
u^\mu_{\rm L} \propto u_{\rm L}^\nu T^{\mu}_{\nu}.
\end{equation}  
In contrast to the Eckart frame, the energy diffusion is absent, $q^{\mu} = 0$.    
We note that nonrelativistic hydrodynamics do not have these ambiguity.
In the nonrelativistic limit,   
the energy current is identical to the particle current because 
the mass energy dominates the energy of fluids.
Actually, the Navier-Stokes equation does not have such ambiguity and the frames.  
It is considered that this difference between the frames is just by the references frames and apparent.
However, several differences, which are not just apparent, actually exist.
For example, the Eckart frame has the instability of the global equilibrium state  at the rest frame, but the Landau frame does not.

To discuss the difference of the Landau and Eckart frames,  we consider fluctuations from the global equilibrium state,
namely, the linear nonequilibrium regime.
The merit of this fluctuating state is that we can observe the state at the same rest frame for the energy and particle currents.
We note that, at the equilibrium state, the particle and energy currents rest: 
$u^\mu_{\rm L} = u^\mu_{\rm E} = (1, \bm{0})$.
Then, in the fluctuating state, 
we also have the same reference frame for the Landau and Eckart frames because
the considered state is just perturbed from the equilibrium one; 
moreover, we need not to bother about what are { \it local equilibrium } and {\it local rest } 
for the relativistic system.
Therefore, in this paper, we focus on the linear fluctuations from the thermal equilibrium state at the rest frame.   

To see relativistic hydrodynamics in the linear regime, 
let us consider the Landau and Eckart equations as examples. 
For the Landau equation, the dissipative terms read
 \begin{align}
\nu^{\mu} &= \lambda \bl( \frac{n T}{h} \br)^2 \partial_{\perp}^\mu (\beta \mu ), \label{eq: d1} \\
q^{\mu} &=  0, \\
\tau^{\mu \nu}&=\eta \bl[ \partial^{\mu}_{\perp}u^{\nu}
               +\partial^{\nu}_{\perp}u^{\mu}
               -\frac{2}{3}\Delta^{\mu\nu}(\partial_{\perp}{\cdot}u)\br] 
               +\zeta\Delta^{\mu\nu}(\partial_{\perp}{\cdot}u), \label{eq: d2}                         
\end{align} 
where $\lambda$, $\eta$ and $\zeta$ are the thermal conductivity, the  share and 
 bulk viscosities, respectively.
 $\Delta^{\mu\nu} \equiv g^{\mu \nu}-u^{\mu} u^{\nu}$ is a projection 
and $\partial^{\mu}_{\perp} \equiv \Delta^{\mu \nu}\partial_{\nu}$ is  the space-like derivative.

Now, we linearize the Landau equation about fluctuations from the equilibrium state.
Let us write 
$n(x)=n_0+\delta n(x)$, 
$e(x)=e_0 + \delta e (x)$, $P(x)=P_0+\delta P(x)$, 
$(\beta \mu )(x)=(\beta\mu)_0 + \delta (\beta\mu) (x)$, and $u^\mu (x)=u^\mu_0 + \delta u^\mu (x)$.
Here, the symbols with the prefix $\delta$, denote the fluctuations.
The equilibrium values are denoted by the suffix $0$.
Hereafter, we employ variables with the suffix and the prefix as the
equilibrium values and fluctuations, respectively.
 For simplicity, let us choose the rest frame as the reference frame:
$u_0^\mu=(1,\bm{0})$.
Then, by the relation in the linear regime, $u_0^{\mu}\delta u_{\mu}=0$,
the fluid-velocity fluctuation is written as
\begin{equation}
\delta u^{\mu }= (0, \delta \bm{v}_{\rm L} ).
\end{equation}   
In consequence, the Landau equation is linearized as
\begin{align}
\partial_0 \delta n &=-n_0 \bm{\nabla} \cdot \delta \bm{v}_{\rm L} 
  +\lambda \bl( \frac{n_0 T_0}{h_0} \br)^2 \bm{\nabla}^2 \delta ( \beta\mu ), \label{eq: ln}\\
\partial_0 \delta e&= - h_0\bm{\nabla} \cdot \delta \bm{v}_{\rm L} , \label{eq: le}\\
\partial_0 (h_0 \delta \bm{v}_{\rm L})&= -\bm{\nabla}(\delta P)
  +\bl( \zeta +\frac{1}{3}\eta \br) \bm{\nabla}(\bm{\nabla}\cdot\delta\bm{v}_{\rm L} ) 
  +\eta \bm{\nabla}^{2}\delta\bm{v}_{\rm L} .   \label{eq: lj}
\end{align}  
Next, let us move on to the Eckart equation.
the dissipative terms of the Eckart equation are
\begin{align}
\nu^{\mu}&= 0, \\
q^{\mu} &= \lambda ( \partial _{\perp}^{\mu}T -T D u^\mu ), \\
\tau^{\mu\nu}&= \eta \bl[ \partial^{\mu}_{\perp}u^{\nu}+
   \partial^{\nu}_{\perp}u^{\mu}-\frac{2}{3}\Delta^{\mu\nu}(\partial_{\perp}{\cdot}u) \br] 
  +\zeta\Delta^{\mu\nu}(\partial_{\perp}{\cdot}u),
   \label{eq:eck}
\end{align}
where $D \equiv u^{\mu}  \partial_{\mu}$ is the time-like derivative.
The linearized equations are  
\begin{align}
\partial_0 \delta n&=-n_0 \bm{\nabla} \cdot \delta \bm{v}_{\rm E}, \label{eq: eckn}\\
\partial_0 \delta e&= - h_0\bm{\nabla} \cdot \delta \bm{v}_{\rm E}
+\lambda (\bm{\nabla}^2 \delta T+T_0\partial_0\bm{\nabla} \cdot \delta {\bm{v}}_{\rm E}), \label{eq: ecke}\\
\partial_0 (h_0\delta \bm{v}_{\rm E})
& = -\bm{\nabla}(\delta P)+\eta \bm{\nabla}^{2}\delta\bm{v}_{\rm E}
+\bl( \zeta +\frac{1}{3}\eta \br) \bm{\nabla}(\bm{\nabla}\cdot\delta\bm{v}_{\rm E} )  
 + \lambda \bl( \bm{\nabla} \partial_0 \delta T+ T_0 \partial_0^2\delta \bm{v}_{\rm E} \br) .  \label{eq: eckj}
\end{align}
We note that the linearized Landau and Eckart equations have different forms even in the same rest frame, as we mentioned.

To investigate relativistic hydrodynamics in the linear regime,
we here use Mori's projection operator method \cite{mori}.
Mori's projection operator method is a powerful tool for extracting slow dynamics.
This method is widely applied and succeed in condensed matter physics \cite{onuki,mazenko,dynamical}.
Actually, various slow dynamics, e.g. the Navier-Stokes, Langevin, Boltzmann equations, and equations for Nambu-Goldstone bosons, are derived
\cite{moriboltzmann,mazenko,Hidaka:2012ym}.
The merit of the projection operator method is that we can derive slow dynamics 
only by choosing slow variables and commutation relations of those without microscopic details.
We note that dynamics on macroscopic scale can be described by much fewer degrees of freedom
than those on microscopic scale.
Such degrees of freedom are called the slow variables (or gross variables). 
The slow variables are degrees of freedom 
that label a macroscopic state and describe long-time behavior.

The view point of the projection operator method tells us 
that a choice of the slow variables is a key for hydrodynamics.
From the linearized Landau and Eckart equations, 
(\ref{eq: ln})-(\ref{eq: lj}) and (\ref{eq: eckn})-(\ref{eq: eckj}),
we see that the dynamic variables are given as the energy and particle densities,
and fluid-velocity fluctuations
\footnote{We note that the intensive variables, like $\delta (\beta \mu) and \delta T$, 
 turn out to be functions of $\delta n, \delta e$ by thermodynamic relations.}:
\begin{equation}
\{\delta e, \delta n,  \delta v^i_{\rm L ,E}\}.
\end{equation} 
The fluid velocities are different depending on the frames:
\begin{align}
\delta v_{\rm E}^i &= n_0^{-1} j^{i},\\
\delta v_{\rm L}^i &= h^{-1}_0 T^{0 i}.
\end{align}
Namely, the difference of the frames is the choices of the slow variables.       

The important point about the choices is that $j^{i}$ is not essentially slow
because it is not a conserved charge.
In contrast, $T^{0 i}$ is a conserved charge and slow because 
the energy current is equivalent to the momentum for the relativistic system: $T^{0i} = T^{i0}$.
Actually, we shall find that a slow part in $j^{i}$ originates from $T^{0i}$.
We will provide detail on this point in Sec.~\ref{sec:slowvariable} and \ref{sec:onsager}.
Then, in this paper, we will apply the projection operator method for the choices,
\begin{align}
 &\{\delta e, \delta p^i, \delta n \}, \label{eq:lset} \\
 &\{ \delta e, \delta p^i,  \delta n,\delta j^i \}, \label{eq:eset} 
\end{align}
where $\delta p^{i} \equiv T^{0 i}$ and $\delta j^{i} \equiv j^{i}$.
We will show that the first set of slow variables, Eq.~(\ref{eq:lset}), 
gives the linearized Landau equations~(\ref{eq: ln})-(\ref{eq: lj}).
Furthermore, we will derive the equations for the second set, Eq.~(\ref{eq:eset}),
which include the Landau and Eckart frames.
Then, we will derive the linearized Eckart equations by eliminating $\delta p^i$
from the equations for $\{\delta e, \delta p^i, \delta n, \delta j^i \}$. 
We note that, to correctly treat the slow part of $\delta j^i$ coming from $\delta p^i$,
we have
 to choose the both of them as the slow variables. 

 After that, we will discuss that the Landau frame is natural for the slow dynamics.
In particular, we shall show that the equations for  $\{\delta e, \delta p^i, \delta n, \delta j^i \}$,
which include the both of the Landau and Eckart frame, have the same slow modes as the Landau equation has.
Namely, the slow modes are determined only by  $\{\delta e, \delta p^i, \delta n \}$, and
the current density $\delta j^i$ contains an irrelevant part for the slow dynamics. 
Moreover, we will illustrate that, even for the Eckart equation, 
the slow dynamics is actually described by the Landau's variables,
 $\{\delta e, \delta p^i, \delta n \}$. 

Our study is first attempt for the derivation based on the projection operator method.
The earlier studies \cite{Denicol,Van,Muronga,Tsumura:2011cj} assume the relativistic Boltzmann equation as an underlying microscopic theory.
In contrast, we stress that our derivation is independent of microscopic details.

This paper is organized as follows.
In Sec.~\ref{sec:POM}, we briefly review Mori's projection operator method for readers being unfamiliar with it.
Also, we explain conserved charges as the slow variables and 
discuss those for relativistic hydrodynamics. 
In Sec.~\ref{sec:metric}, we first determine equal-time correlations of the slow variables
to determine the property of the equilibrium state.
We note that we here consider the fluctuations from {\it the equilibrium state}.
Our determination is based on thermodynamics and Lorentz symmetry on microscopic scale.
Then, it is independent of the microscopic detail.
In Sec.~\ref{sec:application},
 we derive slow dynamics by the projection operator method for the sets, 
Eqs.~(\ref{eq:lset}) and (\ref{eq:eset}).
 In Sec.~\ref{sec:onsager}, we discuss that the Landau frame is natural for relativistic hydrodynamics.
In particular, we study slow modes of the equations for $\{\delta e, \delta p^i, \delta n \}$ 
and $\{\delta e, \delta p^i, \delta n, \delta j^i \}$.
We explicitly show that these equations have the same modes.
Furthermore,
we consider the Onsager reciprocal relation in the Eckart equation to 
illustrate that the slow part of $\delta \bm{j}$ coming from
$\{\delta e, \delta p^i, \delta n \}$.

\section{Mori's projection operator method}
\label{sec:POM}
In this section, we provide Mori's projection operator method \cite{mori,zwanzig1,zwanzig2,mazenko,dynamical}.
We can formally extract slow dynamics from microscopic Hamiltonian dynamics by this method.
On the microscopic scale, an operator at time $t$, $\hO(t)=e^{i\hat{H}t}\hO(0)e^{-i\hat{H}t}$, evolves
by the Heisenberg equation, 
\begin{equation}
\partial_0 \hO (t) = i[\hat{H},\hO (t)] \equiv i\Lv\hO (t),
\label{eq:HeisenbergEquation}
\end{equation}
where $\Lv$ is the Liouville operator.
In the following, we decompose this time evolution equation into  slow and fast ones.

\subsection{Projection operator}
First, we introduce basic ingredients for the projection operator.
Let us consider a many-body system at finite temperature.
As an equilibrium distribution, we assume the grand-canonical one.
Then, the density matrix is given as 
\begin{equation}
\hat{\rho}_\text{eq} \equiv\frac{e^{-\beta (\hH -\mu \hN )}}{\tr e^{-\beta (\hH -\mu \hN)}},
\end{equation}
where $\hH$ is the Hamiltonian, $\hN$ is the number operator,  $\mu$ the chemical potential,
and the inverse temperature $\beta=1/T$. 
With the density matrix, thermal average of $\hO (t)$ is defined as
\begin{equation}
\begin{split}
\average{\hO (t)}
\equiv\tr \hat{\rho}_\text{eq}\hO (t) =\average{\hO (0)}.
\end{split}
\end{equation}
Also, we define an inner product of $\hA$ and $\hB$ as
\begin{equation}
(\hA,\hB)  \equiv \frac{1}{\beta}\int_0^\beta d\tau \average{e^{\tau (\hH -\mu \hN)}\hA e^{-\tau (\hH-\mu \hN)} \hB^\dag}
= \frac{1}{\beta}\int_0^\beta d\tau \average{ \hA(-i\tau) \hB^\dag}.
\label{eq:innerProduct}
\end{equation}

Moreover, we introduce a set of slowly-varying operators (slow variables),
$\{ \hA_n (t, \bm{x})\} = \{ \hA_1, \hA_2, ... ,\hA_n \}$.
If we can separate the time scale into long- and short-time ones, 
such operators exist and describe the slow dynamics.
Let us consider the slow operators at the initial time $t=0$, $\{ \hA_n(0,\bm{x}) \}$.
In general, they are not orthogonal to each other.
We  introduce metric to consider the orthogonal basis:
\begin{equation}
g_{nm}(\bm{x}-\bm{y})   \equiv (\hA_n(0,\bm{x}), \hA_m(0,\bm{y})).
\end{equation}
The orthogonal operators, represented with an upper index, is defined as
\begin{equation}
\hA^{n}(t,\bm{x}) \equiv\int d^3y g^{nm}(\bm{x}-\bm{y})\hA_{m}(t,\bm{y}),
\end{equation} 
where $g^{nm}(\bm{x} - \bm{y})$ is the inverse of $g_{nm}(\bm{x}-\bm{y})$.
These quantities are orthogonal to those with lower indices,
\begin{align}
(\hA_n(0,\bm{x}), \hA^{m}(0,\bm{y})) &= \delta_{n}^{~m} \delta(\bm{x}-\bm{y}), \\
\sum_m\int d^3yg_{nm}(\bm{x}-\bm{y})g^{ml}(\bm{y}-\bm{z}) &= \delta_{n}^{~l}\delta(\bm{x}-\bm{y}).
\end{align}

We have prepared the basic ingredients.
Let us introduce the projection operator acting on any operators $ \hB (t, \bm{x})$ as
\begin{equation}
\cP \hB(t,\bm{x}) \equiv \sum_n\int d^3y  \hA_n(0,\bm{y}) (\hB (t,\bm{y}),\hA^n (0,\bm{x})).
\label{eq:projector}
\end{equation}  
The projection operator extracts the slowly varying part of $\hB$, 
which is determined only by the slow variables $\{ \hA_n \}$. 
We also define the orthogonal projector as $\cQ \equiv 1-\cP$ for later use.

\subsection{Generalized Langevin equation}
In this subsection, we derive so-called the generalized Langevin equation. 
This equation is given by decomposing the Heisenberg equation into slow and fast parts.
For the decomposition, we use the following operator identity:
\begin{equation}
\partial_0 e^{i \Lv t}=e^{i \Lv t} \cP i \Lv
+\int^t_0ds e^{i\Lv (t-s)} \cP i \Lv  e^{ \cQ i \Lv s} \cQ i \Lv
+e^{ \cQ i \Lv t} \cQ i \Lv, \label{eq:operatoridentity}
\end{equation}
which is valid for arbitrary $\Lv$ and $ \cP $ \cite{onuki}.

Let us derive this identity. First,  we consider the following decomposition:
\begin{align}
\partial_0 e^{i \Lv t} &= e^{i \Lv t} i\Lv 
                                         = e^{i \Lv t} \cP i\Lv +e^{i \Lv t}  \cQ i\Lv. \label{eq:oi2}
\end{align}
Next, we consider the Laplace transform of $\exp({i \Lv t})$,
\begin{equation}
 \int_{0}^{\infty} dt e^{-z t} e^{i \Lv t} = \frac{1}{z - i \Lv }.
 \label{eq:LaplaceTransform}
\end{equation}
Then, we decompose Eq.~(\ref{eq:LaplaceTransform}) into
\begin{align}
\frac{1}{z - i \Lv } &= \frac{1}{z - i \Lv }(z - \cQ i\Lv)\frac{1}{z - \cQ i\Lv} \notag \\
                     &= \frac{1}{z - i \Lv }(z - i\Lv + \cP i \Lv)\frac{1}{z - \cQ i\Lv} \notag \\
                     &= \frac{1}{z -\cQ i\Lv } + \frac{1}{z - i \Lv} \cP i\Lv \frac{1}{z -\cQ i\Lv }.
\end{align}
Performing the inverse Laplace transform, we find the identity,
\begin{equation}
e^{i \Lv t}=e^{\cQ i \Lv t}+\int_{0}^{t}ds e^{i \Lv (t-s)} P i\Lv e^{\cQ i \Lv s} \label{eq:oi3}.
\end{equation}
Substituting Eq.~(\ref{eq:oi3}) into the second term of Eq.~(\ref{eq:oi2}), 
we obtain the operator identity, Eq.~(\ref{eq:operatoridentity}).

Multiplying  Eq.~(\ref{eq:operatoridentity}) by an initial value of the slow operator, 
we obtain the decomposed equation of motion for $\hA_n (t) = e^{i \Lv t} \hA_n (0)$:
\begin{equation}
\partial_0 \hA_n(t,\bm{x})=\int d^3y i{\varOmega_n}^m(\bm{x}-\bm{y})\hA_m(t,\bm{y})
-\int_0^\infty dsd^3y {\memory_n}^m(t-s,\bm{x}-\bm{y}) \hA_m(s,\bm{y})+\noise_n(t,\bm{x}),
\label{eq:moriProjectioin}
\end{equation}
without any approximations.
Here, we introduced the following functions and operator:
\begin{align}
i{\varOmega_{n}}^m(\bm{x}-\bm{y})&\equiv(i\Lv \hA_n(0,\bm{x}),\hA^m(0,\bm{y}))=-\frac{1}{\beta}i\average{ [\hA_n(0,\bm{x}),\hA^{m\dag}(0,\bm{y})]} , \label{eq:streaming}\\
{\memory_n}^m(t-s,\bm{x}-\bm{y})&\equiv-\theta (t-s)(i\Lv \noise_n(t-s,\bm{x}), \hA^m(0,\bm{y})),
\label{eq:memo}\\
\noise_n (t, \bm{x})&\equiv e^{it\cQ \Lv}\cQ i\Lv \hA_n (0, \bm{x}), \label{eq:Rn}
\end{align}
 where $\theta (t-s) $ in Eq.~(\ref{eq:memo}) is the Heaviside step function.

Equation (\ref{eq:moriProjectioin}) is the generalized Langevin equation and 
has the following properties:
\begin{enumerate}
\item Equation~(\ref{eq:moriProjectioin}) is the {\em operator identity}. 
\item The first and second terms in the right-hand side represent the slow motions.
\item The first term corresponds to a time-reversible change.
\item The second term corresponds to a time-irreversible change.
      Also, this term depends on a past time value, $\hA_m(s)$, for $s < t$.
         ${\memory_n}^m(t-s,\bm{x}-\bm{y})$ is called the memory function.
\item The last term is the noise term corresponding to the fast motion. 
      For hydrodynamics, this term is usually neglected.
      On the other hand, for the Langevin dynamics, we treat this term as a random noise.
\end{enumerate}

It is useful to rewrite Eq.~(\ref{eq:moriProjectioin}) to the equation in momentum space:
\begin{equation}
\partial_0 \hA_n(t,\bm{k})= i{\varOmega_n}^m(\bm{k})\hA_m(t,\bm{k})
-\int_0^\infty ds {\memory_n}^m(t-s,\bm{k}) \hA_m(s,\bm{k})+\noise_n(t,\bm{k}).
\label{eq:moriProjectioinMomentum}
\end{equation}
For time component, we perform the Laplace transform:
\begin{equation}
\hA_n(z,\bm{k})= \int dt\int d^3x e^{-z t}e^{-i\bm{k}\cdot\bm{x}} \hA_n(t,\bm{x}).
\end{equation}
Then, Eq.~(\ref{eq:moriProjectioinMomentum}) becomes
\begin{equation}
z \hA (z,\bm{k})=  i{\varOmega}(\bm{k}) \hA (z,\bm{k})
- {\memory}(z,\bm{k}) \hA( z,\bm{k})+\noise(z,\bm{k}) +\hA (t=0,\bm{k})
\label{eq:moriProjectioinLaplace}
\end{equation}
in the Laplace-momentum space. Here, $\hA (t=0,\bm{k})$ is the initial value and we used matrix notation.

\subsection{Conserved charges as slow variables}
\label{sec:slowvariable}
Here, we provide why conserved charges are slow 
and discuss the dynamic variables for the Landau and Eckart frames.
The key is that conserved charge densities generally satisfy conservation laws:
\begin{equation}
\partial_0 \hj^0 = -\partial_i \hj^i,
\end{equation} 
where $\hj^0$ is a conserved charged density, and $\hj^i$ is its current.
In the momentum space, the conservation law becomes
\begin{equation}
\partial_0 \hj^0 = i k_i \hj^i.
\end{equation} 
We note that the time change rate of $\hj^0$ is proportional to the wavenumber,
so that the low-wavenumber components turn out to be slow.
Therefore, the change of the conserved charge densities are necessarily slow in the low-wavenumber region, i.e., on macroscopic scale. 

Now, let us consider the case of relativistic hydrodynamics.
We here have the three conservation laws in Eqs.~(\ref{eq:conservation1}) and (\ref{eq:conservation2}).
From those, we obtain the three conserved charges, the particle number, the energy and the momentum:
\begin{align}
\partial_0 \hj^0 &= i k_i \hj^i, \\
\partial_0 \hT^{00} &= i k_i \hT^{0i},\\
\partial_0 \hT^{i0} &= i k_j \hT^{ij}.
\end{align}  
We note that the above quantities are the slow variables for the Landau equation.
The important point is that the particle current is not conserved:
\begin{equation}
\partial_0 \hj^i \neq i k_j \hat{\Pi}^{ij}.
\end{equation}
Thus, the time change rate is not proportional to the wavenumber.
Namely, the particle current is not essentially slow 
although it is proportional to the fluid velocity for the Eckart frame.

Nevertheless, we note that the particle current has a slow part coming from the conserved charges,
because $\hj^i$ is not orthogonal to $\{\hj^0, \hT^{00}, \hT^{0i} \}$.
In other words, the projection of $\hj^i$ on those does not vanish,
\begin{equation}
\cP \hj^i \neq 0,
\end{equation}
and gives the slow part.
From this slow part, we can derive the linearized Eckart equation as we will show in Sec.~\ref{sec:application}.
Here, we stress that the slow dynamics is essentially determined by the conserved charges, $\{\hj^0, \hT^{00}, \hT^{0i} \}$, even for the Eckart equation.

\section{Metric and thermodynamic quantities}
\label{sec:metric}
In this section, we discuss relations between the metric $g_{nm}$ and thermodynamic quantities~\cite{onuki,mazenko,Giusti:2010bb}.
As discussed in Sec.~\ref{sec:intro}, we employ the fluctuations of conserved charges as slow variables, i.e., $\hA_n=\{\dele, \delp^i, \deln, \delj^i\}$
with $\dele\equiv \hT^{00}-e_0$, $\delp^i\equiv \hT^{0i}$, $\deln\equiv \hj^0-n_0$, and $\delj^i\equiv \hj^i$ with $e_0\equiv\average{\hT^{00}}$ and $n_0\equiv\average{\hj^0}$.
We assume that the density matrix at thermal equilibrium is  invariant under time reversal transformation, i.e., $\mathcal{T}\hat{\rho}_\text{eq}\mathcal{T}^{-1} =\hat{\rho}_\text{eq}$, where $\mathcal{T}$ is the time reversal operator.
The slow variables transform under $\mathcal{T}$ as
\begin{align}
\mathcal{T}\dele(t,\bm{x})\mathcal{T}^{-1}&=\dele(-t,\bm{x}),\qquad \
\mathcal{T}\deln(t,\bm{x})\mathcal{T}^{-1}=\deln(-t,\bm{x}), \label{eq:tr1} \\
\mathcal{T}\delp^i(t,\bm{x})\mathcal{T}^{-1}&=-\delp^i(-t,\bm{x}),\quad
\mathcal{T}\delj^i(t,\bm{x})\mathcal{T}^{-1}=-\delj^i(-t,\bm{x}). \label{eq:tr2}
\end{align}
$\dele$ and $\deln$ ($\delp^i$ and $\delj^i$) are even (odd) operators, so that  $\deln$ and $\dele$ does not mix $\delp^i$ and $\delj^i$, i.e., $g_{ep}(\bm{k})=g_{ej}(\bm{k})=g_{np}(\bm{k})=g_{nj}(\bm{k})=0$.

Since we are interested in the low energy behavior of slow variables, we apply the derivative expansion.  The metric is expanded as  a power series of $k^i$,
\begin{equation}
g_{nm}(\bm{k})= g_{nm} +  g^{(1)}_{nm;i}k^i+g^{(2)}_{nm;ij}k^ik^j + \cdots,
\end{equation}
where we assumed $g_{nm}(\bm{k})$ is analytic at $\bm{k}=\bm{0}$; in other words, there are no long range correlations.
The only leading terms, $g_{nm}$, contribute to the linearized hydrodynamic equations at first order, so that we will not consider contributions from the higher order terms.

\subsection{$g_{nn}$, $g_{ee}$ and $g_{en}$}
 First, we focus on $g_{nn}$, $g_{ee}$ and $g_{en}$, which are nothing but susceptibilities,
\begin{align}
g_{ee}=&\int d^3x(\dele(\bm{x}),\dele(\bm{0})) = \frac{1}{V}\average{(\delta \hH)^2},\\
g_{nn}=&\int d^3x(\deln(\bm{x}),\deln(\bm{0}))=  \frac{1}{V}\average{(\delta \hN)^2}, \\
g_{en}=&g_{ne}=\int d^3x(\dele(\bm{x}),\deln(\bm{0}))=  \frac{1}{V}\average{\delta \hH \delta \hN },
\end{align}
where $V$ is the volume, $\delta \hH=\hH -\average{\hH}$ and $\delta \hN=\hN- \average{\hN}$.
We also used the following relations:
\begin{align}
\int d^3x e^{\tau (\hH -\mu \hN )}\dele (\bm{x}) e^{-\tau (\hH -\mu\hN )}&=\int d^3x \dele(\bm{x})=\delta \hH,\\
\int d^3x e^{\tau (\hH -\mu \hN )}\deln (\bm{x}) e^{-\tau (\hH -\mu\hN )}&=\int d^3x \deln(\bm{x})=\delta \hN.
\end{align}
 Using the grand partition function $Z\equiv\tr \exp(-\beta(\hH -\mu \hN ))$, we can rewrite these susceptibilities as
\begin{align}
g_{ee} =& \frac{1}{V} \bl( \frac{\partial^2 \ln{Z}}{\partial \beta^2 }   \br)_{\beta\mu}=-\bl( \frac{\partial e}{\partial \beta} \br)_{\beta\mu, V}, \label{eq:gee} \\
g_{nn} =&\frac{1}{V} \bl( \frac{\partial^2 \ln{Z}}{\partial (\beta \mu)^2 }   \br)_{\beta}= \bl( \frac{\partial n}{\partial (\beta\mu)} \br)_{\beta, V}, \label{eq:gnn}\\
g_{en} =& \frac{1}{V} \bl( \frac{\partial^2 \ln{Z}}{\partial \beta \partial (\beta\mu)}   \br)=\bl( \frac{\partial e}{\partial (\beta\mu) } \br)_{\beta , V} = - \bl( \frac{\partial n}{\partial \beta} \br)_{\beta\mu, V}. \label{eq:gen} 
\end{align}
 
From these equation, the inverse matrices for $e$ and $n$ are obtained as 
\begin{align}
g^{ee}=&-\bl( \frac{\partial \beta}{\partial e}\br)_n, \\
g^{nn}=&\bl(\frac{\partial (\beta\mu)}{\partial n}\br)_e, \\
g^{ne}=&g^{en}=-\bl(\frac{\partial \beta }{\partial n}\br)_e =\bl(\frac{\partial (\beta\mu) }{\partial e}\br)_n .
\end{align}
Then, the slow variables with upper index are 
\begin{align}
\hA^e&=-\left( \frac{\partial \beta}{\partial e}\right)_n \dele
-\left( \frac{\partial \beta}{\partial n}\right)_e \deln \equiv -\delbeta,\\
\hA^n&=\left( \frac{\partial (\beta \mu)}{\partial e}\right)_n \dele
+\left( \frac{\partial(\beta \mu)}{\partial n}\right)_e \deln \equiv \delmu,
\end{align}
 where we introduced the operators for the fluctuations of $\beta$ and $\beta\mu$.
Namely, those are orthonormal to $\dele$ and $\deln$ in the leading order of the derivative expansion:
\begin{align}
(\dele (t, \bm{r}), \delmu (t, \bm{r}') ) &=
 (\deln (t, \bm{r}), \delbeta (t, \bm{r}'))=0, \label{eq:ort1}\\
-(\dele (t, \bm{r}), \delbeta (t, \bm{r}'))&=
(\deln (t, \bm{r}), \delmu ) (t, \bm{r}'))= \delta (\bm{r}-\bm{r}')
\label{eq:ort2}.
\end{align} 

For later use, we also introduce the operators for  the pressure and the temperature fluctuations as
\begin{align}
\delP &\equiv \left( \frac{\partial P}{\partial e}\right)_n \dele
+\left( \frac{\partial P}{\partial n}\right)_e \deln ,\\
\delT &\equiv \left( \frac{\partial T }{\partial e}\right)_n \dele
+\left( \frac{\partial T}{\partial n}\right)_e \deln .
\end{align}
Those operators satisfy the usual thermodynamic relations, e.g, the Gibbs-Duhem relation:
\begin{align}
\delP &  =\frac{h_0}{T_0}\delta \hT +T_0 n_0 \delmu ,   \label{eq:GD}
\end{align}
because we can apply the thermodynamic relations to the coefficients, like $ ( \partial P / \partial e )_n $, in the definitions.

\subsection{$g_{p^i p^j}$, $g_{p^ij^j}$, and $g_{j^ij^j}$}\label{sec:Poincare}
Next, we consider $g_{p^i p^j}$, $g_{p^ij^j}$ and $g_{j^ij^j}$.  
We show that two of them, $g_{p^i p^j}$ and $g_{p^ij^j}$, are expressed as 
the enthalpy and the number density:
\begin{align}
g_{p^i p^j} &= \int d^3x(\hT^{0i}(0,\bm{x}),\hT^{0j}(0,\bm{0}))
= \delta^{ij}T_0h_0, \label{eq:gpp}\\
g_{p^ij^j} &= \int d^3x(\hT^{0i}(0,\bm{x}),\hj^{i}(0,\bm{0}))
= \delta^{ij}T_0n_0. \label{eq:gpj}
\end{align}
The same relations are obtained in Ref.~\cite{Giusti:2010bb}.
These relations are derived from Lorentz symmetry underlying theory.
For an arbitrary Hermitian operator $\hO$,
the following identity is satisfied:
\begin{equation}
\begin{split}
\int d^3x(\hT^{0i}(0,\bm{x}),\hO)= (\hat{P}^i, \hO)
=i ([\hH, \hat{K}^i], \hO )=-iT\average{ [\hat{K}^i,\hO ] },
\label{eq:thermodyamicRelation}
\end{split}
\end{equation}
where $\hat{K}^{i}$ is the boost operator, $[\hH, \hat{K}^i]=-i\hat{P}^i$ and the following Kubo's identity was employed:
\begin{equation}
 ([\hH, \hA], \hB)=-T\average{ [\hA,\hB^\dag] }.
\end{equation}
Since $\hT^{\mu\nu}(x)$ and $\hj^{\mu}(x)$ are Lorentz tensor and vector, respectively, these transform under Lorentz transformation as
\begin{align}
[\hat{L}^{\mu\nu}, \hT^{\lambda\rho}(x)]&=i(x^\mu\partial^\nu-x^\nu\partial^\mu)\hT^{\lambda\rho}(x)
-i(\eta^{\mu\lambda}\hT^{\nu\rho}(x)-\eta^{\nu\lambda}\hT^{\mu\rho}(x)+
\eta^{\mu\rho}\hT^{\lambda\nu}-\eta^{\nu\rho}\hT^{\lambda\mu}(x)),\\
[\hat{L}^{\mu\nu}, \hj^{\rho}(x)]&=i(x^\mu\partial^\nu-x^\nu\partial^\mu)\hj^{\rho}(x)-i(\eta^{\mu\rho}\hj^\nu(x) -\eta^{\nu\rho}\hj^\mu(x)),
\end{align}
where $\hat{L}^{\mu\nu}$ is the charge of Lorentz symmetry, and $\eta^{\mu\nu}=\mathrm{diag}(1,-1,-1,-1)$ is the (inverse) Minkowski metric.
For the Lorentz boost $\hat{K}^i= \hat{L}^{i0}$, they obey
\begin{align}
[\hat{K}^i, \hT^{0j}(x)]&=-i(x^0\partial^i-x^i\partial^0 ) \hT^{0j}(x)+i\hT^{ij}-i\eta^{ij}\hT^{00}(x),\\
[\hat{K}^i, \hj^{j}(x)]&=-i(x^0\partial^i-x^i\partial^0 )\hj^{j}(x) -i\eta^{ij}\hj^0(x).
\end{align}
Therefore, the thermal averages for these commutators satisfy
\begin{align}
\average{ [\hat{K}^i, \hT^{0j}(x)] }&=i\average{ \hT^{ij}(x)-\eta^{ij}\hT^{00}(x)}=i\delta^{ij}h_0, \label{eq:commutationT0i}\\
\average{ [\hat{K}^i, \hj^{j}(x)]}&=-i\average{ \eta^{ij}\hj^0(x)}=i\delta^{ij}n_0. \label{eq:commutationN}
\end{align}
Inserting Eqs.~(\ref{eq:commutationT0i}) and (\ref{eq:commutationN})
into Eq.~(\ref{eq:thermodyamicRelation}), we arrive at Eqs.~(\ref{eq:gpp}) and (\ref{eq:gpj}).
These identities enable us to relate two-point functions to one-point functions.

\section{Application of Mori's projection operator method to relativistic hydrodynamics}
\label{sec:application}
In this section, we apply Mori's projection operator method to relativistic hydrodynamic systems, and derive equations of motion for $\{\dele, \delp^i, \deln \}$ 
and $\{\dele, \delp^i, \deln, \delj^i \}$.
We first show that the set, $\{\dele, \delp^i, \deln \}$, gives the linearized Landau equation.
For the Eckart equations, we introduce the current of the conserved charge, $\delj^i$, which is proportional to the fluid velocity in the Eckart frame, in addition to $\delp^i$ and $\deln$.
We employ the derivative expansion and keep the spatial and time derivative  to the second order, i.e., $\partial_0,\bm{\nabla},\bm{\nabla}^2,\partial_0\bm{\nabla}$ and $\partial_0^2$. We will drop the noise term $\noise_n(t, \bm{x})$ in the equation of motion.
This term is irrelevant in the time evolution of the expectation value. 
If one is interested in stochastic hydrodynamics, one may keep the noise term~\cite{LandauText}.

\subsection{Linearized Landau equation}
First, we derive the linearized Landau equation.
For this purpose, we choose $\dele$, $\delp$ and $\deln$ as slow variables.
Since $\delp^i$ is chosen as a slow variable,
the equation  for $\partial_0\dele$ does not contain dissipative terms,
\begin{equation}
\partial_0 \dele = -\bm{\nabla}\cdot \vdelp = -h_0\bm{\nabla}\cdot \vvL,
\label{eq:LEPOM}
\end{equation}
where we defined the fluid velocity $\vvL\equiv \vdelp/h_0$.
This equation is nothing but energy conservation law, Eq.~(\ref{eq: le}).
This can be confirmed by the the following calculation:
\begin{equation}
\begin{split}
i\varOmega_{e}^{~p}(\bm{k})
&= \int d^3x e^{-i\bm{k}\cdot\bm{x}}(i\Lv\dele(\bm{x}),\delp^j(\bm{0}))g^{p^jp^i}(\bm{k})\\
&= \int d^3x e^{-i\bm{k}\cdot\bm{x}}(-{\nabla}^j \delp^j(\bm{x}),\delp^i(\bm{0}))g^{p^jp^i}(\bm{k})\\
&=-ik^l g_{p^lp^j}(\bm{k})g^{p^jp^i}(\bm{k})\\
&=-ik^i.
\end{split}
\end{equation}
Therefore, the reversible term becomes $-\bm{\nabla}\cdot\vdelp$. 
The memory function vanishes
because $i\Lv \dele$ turns out to be $-i\bm{k}\cdot \vdelp$, and then
$\cQ i\Lv \dele=0$ [see Eqs.~(\ref{eq:memo}) and (\ref{eq:Rn})].

Let us move onto  the equation for $\partial_0\deln$.
For the reversible part,  the only $i\varOmega_{np^i} $ survives in $\varOmega$ from time reversal symmetry,  
which is
\begin{equation}
\begin{split}
i\varOmega_{np^i}(\bm{k})&= \int d^3x e^{-i\bm{k}\cdot\bm{x}}(i\Lv\deln(\bm{x}),\delp^i (\bm{0})) \\
&= \int d^3x e^{-i\bm{k}\cdot\bm{x}}(-{\nabla}^j \delj^j(\bm{x}),\delp^i(\bm{0})) \\
&=-ik^j g_{j^jp^i} +\mathcal{O}(k^3) \\
&=  -ik^i T_0n_0  +\mathcal{O}(k^3).
\end{split}
\end{equation}
For the memory function, we keep it up to of order $k^2$.
$\memory_{ne}$ vanishes as the same as previous due to $\cQ i\Lv \dele=0$.
$\memory_{n p}$ is of order $k^3$ from tensor structure, which can be neglected.
Therefore,  we may only consider $\memory_{nn}$.
Since we are interested in slow dynamics, we also expand the memory function in terms of $z$ in addition to $k$:
\begin{equation}
\begin{split}
\memory_{nn}(z,\bm{k})
&=\int_0^\infty dt\int d^3x e^{-zt}e^{-i\bm{k}\cdot\bm{x}} ( e^{it\cQ \Lv}\cQ i\Lv \deln(0,\bm{x}), i\Lv \deln(0,\bm{0}))\\
&=k^i k^j\int_0^\infty dt\int d^3x e^{-zt}e^{-i\bm{k}\cdot\bm{x}} ( e^{it\cQ\Lv}\cQ \delj^i(0,\bm{x}),    \delj^j(0,\bm{0}))\\
&\simeq\bm{k}^2\int_0^\infty dt\int d^3x  (\delj^i(t,\bm{x}) - \frac{n_0}{h_0}\delp^{i}(t,\bm{x}),   \delj^i(0,\bm{0}) - \frac{n_0}{h_0}\delp^{i}(0,\bm{0}) )\\
&=\bm{k}^2\lambda\left(\frac{n_0T_0}{h_0}\right)^2 \equiv  \bm{k}^2 \tilde{\lambda},
\label{eq:memorynn}
\end{split}
\end{equation}
where $\simeq$ denotes the approximation of order $k^2$ and $z^0$.
The approximation of order $z^0$ corresponds to the Markov approximation,
i.e., in the coordinate space, $\varPhi_{nn}(t,\bm{x})\simeq -\tilde{\lambda}\bm{\nabla}^2\delta(t)\delta^{(3)}(\bm{x})$.
We defined the thermal conductivity $\lambda$ as
\begin{equation}
\lambda \equiv \left(\frac{h_0}{n_0T_0}\right)^2
\int_0^\infty dt\int d^3x  (\delj^i(t,\bm{x}) - \frac{n_0}{h_0}\delp^{i}(t,\bm{x}),   \delj^i(0,\bm{0}) - \frac{n_0}{h_0}\delp^{i}(0,\bm{0})) ,
\end{equation}
and we used 
\begin{equation}
\cQ \delj^i(t,\bm{x})\simeq \delj^i(t,\bm{x}) - \frac{n_0}{h_0}\delp^{i}(t,\bm{x})
\end{equation}
in the leading order of the derivative expansion. We note that the second term is important to remove the contribution of zero mode from $\delj^i$.
As a result, we arrive at the equation for $\partial_0\deln$ as
\begin{equation}
\partial_0 \deln = -n_0\bm{\nabla}\cdot \vvL
+\tilde{\lambda}\bm{\nabla}^2\delmu .
\label{eq:LNPOM}
\end{equation}
This equation coincides with Eq.~(\ref{eq: ln}).

Similarly, for $\delp^i$,
the reversible term $i\varOmega_{pe}=-ik^i T_0 h_0$, $i\varOmega_{pn}=-ik^i T_0 n_0$, so that 
\begin{equation}
i\varOmega_{pe}\hA^e+i\varOmega_{pn}\hA^n =-ik^i (-T_0h_0\delbeta+ T_0n_0\delmu ))=-ik^i \delP ,
\label{eq:reversiblePLandau}
\end{equation}
where we used the Gibbs-Duhem's relation, Eq.~(\ref{eq:GD}).  
For the dissipative terms,
$\memory_{pe}$ vanishes, and 
$\memory_{pn}\sim k^3$ from tensor structure can be neglected as the same as the previous.
Therefore, only $\memory_{pp}$ survives in the leading order, which is evaluated as
\begin{equation}
\begin{split}
\memory_{p^ip^k}(z, \bm{k})
&\simeq k^jk^l\int dt\int d^{3}x (\delta\hT^{ij}(t,\bm{x}) - \delta^{ij}\delP(t,\bm{x}),\delta\hT^{kl}(0,\bm{0}) - \delta^{kl}\delP(0,\bm{0}))\\
&=T_0\Bigl(\zeta+\frac{1}{3}\eta\Bigr)k^ik^k+T_0\eta \bm{k}^2\delta^{i k},
\end{split}
\end{equation}
where $\delta\hT^{kl}(t,\bm{x})\equiv \hT^{kl}(t,\bm{x})-\average{\hT^{kl}(t,\bm{x})}$, and we used the same approximation in Eq.~(\ref{eq:memorynn}).
We also used the projection $\cQ\delta\hT^{ij}(0,\bm{x})\simeq\delta \hT^{ij}(0,\bm{x}) - \delta^{ij}\delP(0,\bm{x}) $, which can be shown by using the relation $\average{\hT^{ij}(0,\bm{x})}=\delta^{ij}P$.
The shear and bulk viscosities are defined by Kubo formula as
\begin{align}
\eta&=\beta_0\int_0^\infty dt\int d^3x(\delta\hT^{12}(t,\bm{x}), \delta\hT^{12}(0,\bm{0})),\\
\zeta-\frac{2}{3}\eta&=\beta_0\int_0^\infty dt\int d^3x(\delta\hT^{11}(t,\bm{x})-\delP(t,\bm{x}), \delta\hT^{22}(0,\bm{0})-\delP(0,\bm{0})).
\label{eq:KuboViscosities}
\end{align}
Noting that $\hA^{p^i}=-\beta_0 \vL ^i=-(\beta_0 /h_0 )\delp^i$, we obtain
\begin{equation}
\partial_0\vdelp = -\bm{\nabla} \delP+\eta \bm{\nabla}^2\vvL+\left(\zeta+\frac{1}{3}\eta\right)\bm{\nabla}(\bm{\nabla}\cdot\vvL),
\label{eq:LJPOM}
\end{equation}
which coincides with Eq.~(\ref{eq: lj}).
We have shown that the linearized Landau equations, Eqs~(\ref{eq:LEPOM}), (\ref{eq:LNPOM}) and  (\ref{eq:LJPOM}), are derived by choosing 
$\dele$, $\deln$, and $\delp^i$ as slow variables.

Before closing this subsection, let us consider the detail of $\memory_{nn}(z,\bm{k})$.
The memory function can be written as
\begin{equation}
\begin{split}
\memory(z,\bm{k})= - (\ddot{\varXi}(z,\bm{k})-i\varOmega(\bm{k})\dot{\varXi}(z,\bm{k}))\frac{1}{1+\dot{\varXi}(z,\bm{k})},
\label{eq:memory}
\end{split}
\end{equation}
where we defined
\begin{align}
{\varXi}(t)&\equiv( \hA_n(t), \hA^m),\\
\dot{\varXi}(t)&\equiv(i\Lv \hA_n(t), \hA^m),\\
\ddot{\varXi}(t)&\equiv((i\Lv)^2 \hA_n(t), \hA^m).
\end{align}
For the derivation of Eq.~(\ref{eq:memory}), see Appendix~\ref{sec:MemoryFunction}.
Since the time derivative of a conserved charge variable is slow, $\dot{\varXi}$ is of order $k$; then,  we can estimate
$1/(1+\dot{\varXi})=1+\mathcal{O}({k})$. And then, $\memory_{nn}(z,\bm{k})$  becomes
\begin{equation}
\begin{split}
\memory_{nn}(z,\bm{k})&\simeq - \left[\ddot{\varXi}(z,\bm{k})-i\varOmega(\bm{k})\dot{\varXi}(z,\bm{k})\right]_{nn}\\
&\simeq  \bm{k}^2\left({\varXi}_{jj}(z,\bm{0})-\frac{n_0}{h_0}{\varXi}_{pj}(z,\bm{0})\right).
\end{split}
\end{equation}
From $\dot{\varXi}=-1+z\varXi$,
\begin{equation}
\begin{split}
\varXi_{pj}(z,\bm{0})=-\frac{1}{z}g_{pj}=-\frac{1}{z}n_0T_0,
\end{split}
\end{equation}
where we used $\dot{\varXi}_{pj}(z,\bm{0})=0$.
Therefore, we can write
\begin{equation}
\begin{split}
\memory_{nn}(z,\bm{k})\simeq  \bm{k}^2\left(
{\varXi}_{jj}(z,\bm{0})-\frac{n_0^2T_0}{zh_0}\right)\equiv\bm{k}^2\tilde{\lambda}(z),
\label{eq:xjj}
\end{split}
\end{equation}
where we defined the frequency-dependent-thermal conductivity $\tilde{\lambda}(z)$.
At $z=0$, $\tilde{\lambda}(0)$ coincide with $\tilde{\lambda}$.
This expression will be used in the next subsection to derive the linearized Eckart equation.

\subsection{Linearized Eckart equation}
Next, we derive the linearized Eckart equation.
The charge does not dissipate in the Eckart equation, which implies that the fluid velocity 
is chosen as $\vE \equiv \delj^i / n_0 $. Therefore, we choose $\delj^i$ as a slow variable in addition to
$\{ \dele , \delp^i , \deln \}$.
In order to derive the Eckart equation,  we first derive the equations of motion for
 $\{ \dele, \delp^i, \deln ,\delj^i \}$;
and then, we remove the degrees of freedom of $\delp^i$.

The equation of time evolution for $\dele$ and $\deln$ are trivial
thanks to the conservation laws,
\begin{align}
\partial_0 \dele &= -\bm{\nabla}\cdot \vdelp,  \label{eq:EEPOM}\\
\partial_0 \deln &= -\bm{\nabla}\cdot \vdelj. \label{eq:ENPOM}
\end{align}
As usual, we calculate the reversible terms:
\begin{align}
i\varOmega_{ep^i}(\bm{k})&= \int d^3x e^{-i\bm{k}\cdot\bm{x}}(i\Lv\dele(\bm{x}),\delp^i (\bm{0}))
=  -ik^i T_0h_0  +\mathcal{O}(k^3),\\
i\varOmega_{ej^i}(\bm{k})&= \int d^3x e^{-i\bm{k}\cdot\bm{x}}(i\Lv \deln (\bm{x}), \delj^i (\bm{0}))
=  -ik^i T_0n_0  +\mathcal{O}(k^3),\\
i\varOmega_{nj^i}(\bm{k})&= \int d^3x e^{-i\bm{k}\cdot\bm{x}}(i\Lv\deln(\bm{x}),\delj^i (\bm{0}))= -ik^j g_{j^jj^i}+\mathcal{O}(k^3).
\end{align}
Here the explicit form of $g_{j^jj^i}$ is not obtained; however, it is irrelevant in the leading order of the fluid equations, as will be seen later.
The reversible term for $\partial_0 \delp^i$ is the same as that of the Landau equation in Eq.~(\ref{eq:reversiblePLandau}).
The reversible term for $\partial_0 \delj^i$ becomes
\begin{equation}
\begin{split}
i\varOmega_{jn}\hA^{n}+i\varOmega_{je}\hA^e =-ik^i \bl( g_{jj}\delmu )- T_0n_0\delbeta \br) .
\end{split}
\end{equation}
Since $\delj^i$ and $\delp^i$ are chosen as slow variables, 
the inverse metric contain mixing terms:
\begin{equation}
\begin{pmatrix}
g^{pp} & \ g^{pj}\\
g^{jp} & \ g^{jj}
\end{pmatrix}
=\frac{1}{g_{pp}g_{jj}-g_{pj}g_{jp}}
\begin{pmatrix}
g_{jj} & \ -g_{pj}\\
-g_{jp} & \ g_{pp}
\end{pmatrix}
=\frac{\beta_0}{h_0\beta_0 g_{jj}-n_0^2}
\begin{pmatrix}
\beta_0 g_{jj} & \ -n_0\\
-n_0 & \ h_0
\end{pmatrix}.
\end{equation}
Then, the conjugate variables for $\delj^i$ and $\delp^i$ are
\begin{align}
\hA^{j^i}&=
\frac{\beta_0 }{h_0\beta_0 g_{jj}-n_0^2}(-n_0\delp^i+h_0 \delj^i)= -n_0 g^{jj}( \vL^i- \vE^i),\\
\hA^{p^i}&
=\frac{\beta_0}{h_0\beta_0 g_{jj}-n^2}(\beta_0 g_{jj} \delp^i -n_0\delj^i)
=\beta_0 \vE^i+\beta_0 g_{jj}h_0\hA^{j^i}
=\beta_0 \vL^i-\frac{n_0}{h_0}\hA^{j^i},
\end{align}
where we defined the fluid velocity in the Eckart frame as
$\vE^i \equiv {\delj^i}/{n_0}$.
The memory functions appearing  in $\partial_0\delp^i$ are $\memory_{pp}$ and $\memory_{pj}$, which are both of order $k^2$.
Here, we assume that $\hA^{j}$ is of order $k$, which will be checked later.
Then, $\memory_{pj}\hA^j$ can be neglected because it is of order $k^3$.
Similarly, $\memory_{pp}\hA^p\simeq -\beta\memory_{pp}\vvL$.
Therefore,
\begin{equation}
\partial_0\vdelp = -\bm{\nabla} \delP +\eta \bm{\nabla}^2\vvL+\left(\zeta+\frac{1}{3}\eta\right)\bm{\nabla}(\bm{\nabla}\cdot\vvL).
\label{eq:eomP}
\end{equation}
This is the same equation as the Landau one.

Finally, let us consider  the equation for $ \partial_0\delj^i$, which is written in the Laplace space:
\begin{equation}
\begin{split}
zn_0 \vE^i &=i\varOmega_{jn}\hA^n+i\varOmega_{je}\hA^e - \memory_{jj}\hA^j-\memory_{jp}\hA^p + n_0\vE^i(t=0)\\
&=- g_{jj} \nabla^i \delmu )- \frac{n_0}{T_0}\nabla^i \delT - \memory_{jj}\hA^j-\memory_{jp} \hA^p + n_0 \vE^i(t=0),
\end{split}
\label{eq:vE}
\end{equation}
where $\vvE(t=0)$ is the initial value of the fluctuation.
The important point is that $\varPhi_{jj}$ is not slow and gives a contribution of order $k^0$.
This equation will give the relation between these fluid velocities.
Since we are interested in the first order hydrodynamic equation, it is enough to take into account the difference of the fluid velocity up to of order $k^1$.
In this order, we can neglect $\memory_{jp}$ because it is of order $k^2$.

Let us, now, estimate $\varPhi_{jj}$. 
From Eq.~(\ref{eq:memory}), we obtain
\begin{equation}
\begin{split}
\memory_{jj}(z,\bm{k})
&=\left[ - (\ddot{\varXi}(z,\bm{k})-i\varOmega(\bm{k})\dot{\varXi}(z,\bm{k}))\frac{1}{1+\dot{\varXi}(z,\bm{k})}\right]_{jj}
\\
&=\left[-z+\frac{1}{\varXi(z,\bm{k})}+i\varOmega(\bm{k})\right]_{jj}\\
&=-zg_{jj}+\left[\frac{1}{\varXi(z,\bm{k})}\right]_{jj},
\end{split}
\end{equation}
where we used $\ddot{\varXi}=z\dot{\varXi}-i\varOmega$,  $\dot{\varXi}=z\varXi-1$, and $i\varOmega_{jj}(\bm{k})=0$.
We may estimate $\memory_{jj}(z,\bm{k})$ at $\bm{k}=\bm{0}$ in the leading order.
First, we consider $\varXi_{ep^i}(z,\bm{k})$.
The energy conservation provides $z\varXi_{ep^i}(z,\bm{k})=-ik^j{\varXi}_{p^jp^i}(z,\bm{k})$.
Thus, at $\bm{k}=\bm{0}$, $\varXi_{ep^i}(z,\bm{0})=0$. 
Similarly, one can show $\varXi_{ej}(z,\bm{0})=\varXi_{np}(z,\bm{0})=\varXi_{nj}(z,\bm{0})=0$.
Therefore,  we may consider the terms with $\delp^i$ or $\delj^i$.
They are estimated at $\bm{k}=\bm{0}$ as
\begin{equation}
\begin{split}
\begin{pmatrix}
\varXi_{pp}(z,\bm{0}) &\quad \varXi_{pj}(z,\bm{0})\\
\varXi_{jp}(z,\bm{0}) & \quad \varXi_{jj}(z,\bm{0})
\end{pmatrix}
=\frac{1}{z}
\begin{pmatrix}
g_{pp} &\quad g_{pj}\\
g_{jp} &\quad z \varXi_{jj}(z,\bm{0})
\end{pmatrix}
=\frac{1}{z}
\begin{pmatrix}
h_0 T_0 &\quad n_0 T_0\\
n_0T_0 & \quad\frac{n_0^2T_0}{h_0}+z\tilde{\lambda}(z)
\end{pmatrix},
\end{split}
\end{equation}
where we used $\dot{\varXi}(z,\bm{0})=-1+z\varXi(z,\bm{0})=0$, and Eq.~(\ref{eq:xjj}).
Taking into account the metric, we find
\begin{equation}
\begin{split}
\varPhi_{jj}(z,\bm{0})
&=\frac{(g_{jj}h_0-n_0^2T_0)^2}{h_0^2\tilde{\lambda}(z)}-\frac{1}{h_0}(hg_{jj}-n_0^2T_0)z 
= \frac{1}{(g^{jj})^2\tilde{\lambda}(z)}-\frac{1}{g^{jj}}z.
\end{split}
\end{equation}
Then, the equation of motion  becomes
\begin{equation}
\begin{split}
n_0z \vvE
&=- g_{jj} \bm{\nabla}\delta(\beta\mu)- \frac{n_0}{T_0}\bm{\nabla}\delta T + \frac{n_0}{g^{jj}\tilde{\lambda}(z)} (\vvL-\vvE)
-  n_0z( \vvL- \vvE)+n_0 \vvE(t=0).
\label{eq:migel}
\end{split}
\end{equation}
From Eq.~(\ref{eq:migel}), we obtain
\begin{equation}
\begin{split}
h_0\vvL &
=h_0\vvE+\frac{h_0}{n_0}g^{jj}\tilde{\lambda}(z)\Bigl(n_0z \vvL 
  + g_{jj} \bm{\nabla}\delta(\beta\mu)+ \frac{n_0}{T_0}\bm{\nabla}\delT -n_0\vvE(t=0)\Bigr) \\
&=h_0\vvE-\lambda(z)(T_0z\vvL+\bm{\nabla}\delT)
+ \lambda(z) {T_0^2}g^{pp} (h_0z \vvL+ \bm{\nabla}\delP) -g^{jj}\frac{n_0^2T_0^2}{h_0}{\lambda}(z)\vvE(t=0).
\label{eq:pi}
\end{split}
\end{equation}
The third term in the last line is estimated as $h_0z \vvL+ \bm{\nabla}\delP =h_0\vvL(t=0)+\mathcal{O}(\bm{\nabla}^2)$ from the equation of motion,  Eq.~(\ref{eq:eomP}). Then
\begin{equation}
\begin{split}
h_0\vvL
&=h_0\vvE-\lambda(z)(T_0z\vvL+\bm{\nabla}\delT )
+ \lambda(z) \Bigl(T_0^2g^{pp}h_0 \vvL(t=0)  -g^{jj}\frac{n_0^2T_0^2}{h_0}\vvE(t=0) \Bigr)\\
&=h_0\vvE-\lambda(z)(T_0 z\vvL+\bm{\nabla}\delT -T_0\vvL(t=0))
-\frac{T_0^2n_0}{h_0} \lambda(z) \hA^j(t=0). \label{eq:eomJ}
\end{split}
\end{equation}
This equation is exact for an arbitrary $z$ in the first order of the derivative expansion of spatial coordinate. 
We note that the equations~(\ref{eq:EEPOM}), (\ref{eq:ENPOM}), (\ref{eq:eomP}) and (\ref{eq:eomJ})
are those of motion for $\{  \dele, \delp^i, \deln, \delj^i \}$.
These equations are first derived in this paper by the projection operator method.
We also note that these include the Landau and Eckart frames 
in the view point of the projection operator method.

Now, let us derive the Eckart equation. 
We assume that there is no conjugate variable for $j^i$ at the initial time, i.e., $\hA^j(t=0)=0$. 
We also assume that the change of the variable is so slow that $z$ expansion is applicable:
\begin{equation}
\begin{split}
h_0\vvL
&=h_0\vvE-\lambda(T_0z\vvL+\bm{\nabla}\delta T-\vvL(t=0)).
\end{split}
\end{equation}
In the coordinate space, this equation becomes
\begin{equation}
\begin{split}
h_0\vvL = h_0\vvE-\lambda(T_0\partial_0\vvL+\bm{\nabla} \delT ).
\end{split}
\label{eq:vvEvvL}
\end{equation}
From this equation, one can confirm $\hA^{j^i}=-n_0g^{jj}(\vL^i-\vE^i)=n_0g^{jj}(T_0\partial_0\vL^i+ik^i \delT )/h_0\sim k$. 
Therefore, the assumption
that $\hA^{j}$ is of order $k$ is consistent for slow dynamics.
Solving Eq.~(\ref{eq:vvEvvL}) for $h_0\vvL$, we obtain
\begin{equation}
\begin{split}
h_0\vvL&=\frac{1}{1+\lambda \frac{T_0}{h_0}\partial_0}(h_0\vvE-\lambda\bm{\nabla}\delT ) \\
&\simeq h_0\vvE-\lambda(T_0\partial_0\vvE+\bm{\nabla}\delT ),
\end{split}
\label{eq:zexpansion}
\end{equation}
where we dropped $\partial_0\bm{\nabla} T$ in the last line because it is higher order.
Inserting Eq.~(\ref{eq:pi}) into Eq.~(\ref{eq:eomP}), we find
\begin{equation}
\partial_0(h_0\vvE-\lambda(T_0\partial_0\vvE+\bm{\nabla} \delT ))
= -\bm{\nabla} \delP +\eta \bm{\nabla}^2\vvE+\left(\zeta+\frac{1}{3}\eta\right)\bm{\nabla}(\bm{\nabla}\cdot\vvE).  \label{eq:eckjo}
\end{equation}
This equation is equal to the linearized Eckart equation, Eq.~(\ref{eq: eckj}).

\section{Discussion}
\label{sec:onsager}
 Here, we discuss the slow dynamics is determined by the Landau's variables $\{  \dele, \delp^i, \deln  \}$ and the Landau frame is natural for the hydrodynamics.

First, we study modes of the equations for $\{  \dele, \delp^i, \deln, \delj^i \}$, which include both of  the Landau and Eckart frames
in the projection operator method.  
We will show that those modes are the same as of  the Landau equation.
Namely,  the slow modes are determined only by the Landau's variables,
and the current density $\delj^i$  contains an irrelevant part  for the slow dynamics.

Furthermore,  we discuss the dynamic variables of the Eckart equation.
The Eckart equation has $\{  \dele, \deln, \delj^i \}$ as the dynamic variables, apparently.
Nevertheless, we shall show the slow part of $\delj^i$ is determined by $\{  \dele, \delp^i, \deln  \}$, actually. 
To illustrate this fact, we consider the Onsager reciprocal relation in the Eckart equation.
If we assume that the time reversal property of  $\delj^i$ is odd, the reciprocal relation seems to be violated.
On the other hand, if we regard  $\delj^i$ is a dependent variable of $\{  \dele, \delp^i, \deln  \}$,
we will find that the relation is satisfied. 
Namely,  $\delj^i$ in the Eckart equation is projected on the Landau' variables and its time reversal property is not odd.

\subsection{Modes of the equations for $\{  \dele, \delp^i, \deln, \delj^i \}$}
 Here, we study modes of  the equations for $\{  \dele, \delp^i, \deln, \delj^i \}$, which has the dynamic variables for both of the Landau and Eckart frames.  
Then, we shall find that those modes are the same of the Landau equation. 

In the Fourier space, the equations for $\{  \dele, \delp^i, \deln, \delj^i \}$, (\ref{eq:EEPOM}), (\ref{eq:ENPOM}), (\ref{eq:eomP}) and (\ref{eq:eomJ}),
 are written as the following matrix form:
\begin{equation}
\begin{split}
M
\begin{pmatrix}
&\dele (\omega, \bm{k}) \\ 
&\deln (\omega, \bm{k}) \\
&\delp_{\parallel } (\omega, \bm{k}) \\ 
&\delj_{\parallel } (\omega, \bm{k}) \\
&\vdelp_{\perp  } (\omega, \bm{k}) \\ 
&\vdelj_{\perp  } (\omega, \bm{k}) \\
\end{pmatrix}
=
\begin{pmatrix}
0 \\
0 \\
0 \\
0 \\
0 \\
0
\end{pmatrix}
\end{split},
\end{equation}
where we decomposed $\vdelp$ and $\vdelj$ into the longitudinal and transverse components:
\begin{align}
\delp_{\parallel }  = \frac{\bm{k}}{|\bm{k}|} \cdot \vdelp ,
\qquad & \vdelp_{\perp  }  = \vdelp -\delp_{\parallel } \frac{\bm{k}}{|\bm{k}|}  , \\
\delj_{\parallel }  = \frac{\bm{k}}{|\bm{k}|} \cdot \vdelj ,
\qquad & \vdelj_{\perp  }  = \vdelj -\delj_{\parallel } \frac{\bm{k}}{|\bm{k}|} .
\end{align} 
Also, we introduced the matrix,
\begin{equation}
\begin{split}
M \equiv
\begin{pmatrix}
-i\omega & 0 & ik & 0 & 0 & 0\\
0 & -i\omega & 0  & ik & 0 & 0\\
ik \alpha_{Pe} & ik \alpha_{Pn} & -i\omega+k^2\varGamma_{\parallel } & 0 & 0 & 0\\
ik ik \alpha_{Te} & ik \alpha_{Tn} & -i\omega+\varGamma_j & -(h_0/n_0) \varGamma_j & 0 & 0\\
0 & 0 & 0 & 0 & -i\omega+k^2(\eta / h_0) & 0 \\
0 & 0 & 0 & 0 & -i\omega+\varGamma_j & -(h_0/n_0)\varGamma_j
\end{pmatrix}
\end{split},
\end{equation}
where we defined
\begin{align}
\alpha_{pe}\equiv \bl(\frac{\partial P}{\partial e} \br)_n , \qquad &
\alpha_{pn}\equiv \bl(\frac{\partial P}{\partial n} \br)_e , \\
\alpha_{Te}\equiv \beta_0 h_0 \bl(\frac{\partial T}{\partial e} \br)_n , \qquad &
\alpha_{Tn}\equiv \beta_0 h_0 \bl(\frac{\partial T}{\partial n} \br)_e  ,\\
\varGamma_{\parallel } = \frac{1}{h_0}\bl(\zeta + \frac{4}{3}\eta \br) , \qquad &
\varGamma_j =\frac{\beta_0 h_0}{\lambda} .
\end{align}
We here neglected the frequency dependence of the thermal conductivity.
We can obtain dispersion relations from $\det  M  =0   $.
Those are given as the following to second order in $k$: 
\begin{align}
\omega & \sim -ik^2 \varGamma_{\rm t}, \label{eq:vdmode}\\
\omega & \sim - ik^2 \varGamma_{\rm s}\pm  k c_s , \label{eq:sdmode}\\
\omega & = -ik^2 (\eta / h_0)\label{eq:tmode},  
\end{align} 
where  we introduced the thermal and sound diffusion constants, and the sound velocity: 
\begin{align}
\varGamma_{\rm t} &= \frac{\lambda}{n_0 c_{P}}, \\
\varGamma_{\rm s} &=\frac{1}{2} \bl[ \varGamma_{\parallel }
+\varGamma_{\rm t} \bl( n_0c_P\bl( \frac{T_0 n_0}{h_0} \br)^2 
\bl( \frac{\partial (\beta\mu)}{\partial n} \br)_{e} -1\br) \br] ,
\label{eq:sounddiffusion} \\
c_s &= \bl( \frac{\partial P}{\partial e} \br)_{s/n}.
\end{align} 
Here, $c_P$ is the specific heat at constant pressure.
From Eqs.~(\ref{eq:vdmode})-(\ref{eq:tmode}),
we see that the equations for $\{  \dele, \delp^i, \deln, \delj^i \}$  have the usual hydrodynamic modes:
the thermal diffusion, sound and viscous diffusion modes.
Moreover, these dispersions are the same as for the Landau equation
 \footnote{The sound diffusion constant, Eq.~(\ref{eq:sounddiffusion}), is apparently different from that in \cite{Minami}.
 In \cite{Minami}, the Landau equation is solved by choosing $\deln $ and $\delT $ as 
 the thermodynamic variables, whereas $\deln$ and $\dele$ are chosen in this paper.
 The difference is due to the choices and apparent. }, 
i.e.,  the slow modes are described only by  $\{  \dele, \delp^i, \deln \}$ and the current density $\delj^i$ is irrelevant for slow dynamics.

\subsection{Independent variables of the Eckart equation}
 Here, we discuss that, even for the Eckart equation, the slow part of $\delj^i$ is
determined by the Landau's variables  $\{  \dele, \delp^i, \deln \}$.
 For this purpose, we consider the Onsager reciprocal relation in the Eckart equation.
We will see that the relation is not satisfied if $\delj^i$ is assumed as an independent slow variable,
while it is satisfied if $\delj^i$ is a dependent variable of  $\{  \dele, \delp^i, \deln \}$. 

Now, we  consider the correlations, $\bigl( \partial_0\dele, \deln \bigr)$ 
and $\bigl( \partial_0\deln,\dele \bigr)$.
These correlations must satisfy the relation 
\begin{equation}
\bigl( \partial_0\dele, \deln \bigr) = \bigl( \partial_0\deln,\dele \bigr) ,\label{eq:onsager}
\end{equation}
which comes from the time reversal properties of $\deln$, $\dele$ and the equilibrium state.
We note that this relation is equivalent to the Onsager reciprocal relation in the linear regime \cite{onsager1,onsager2,prigogine}.  
From the Eckart equations, Eqs.~(\ref{eq: eckn})-(\ref{eq: eckj}),  the correlations are written as
\begin{align}
\bigl( \partial_0\deln (t, \bm{r}),\dele(t, \bm{r}') \bigr) 
= &-n_0 \bm{\nabla} \cdot \bigl( \vvE(t, \bm{r}) , \dele (t, \bm{r}') \bigr), \label{eq:dne}\\
\bigl( \partial_0\dele(t, \bm{r}), \deln(t, \bm{r}') \bigr) 
= &- h_0\bm{\nabla} \cdot \bigl( \vvE(t, \bm{r})  ,\deln(t, \bm{r}')\bigr) \nonumber \\
&+\lambda \bl( \bm{\nabla}^2 \bigl( \delT (t, \bm{r}) , \deln(t, \bm{r}')\bigr) 
  +T_0 \bm{\nabla} \cdot\bigl( \partial_0 \vvE(t, \bm{r}),  \deln(t, \bm{r}')\bigr) \br).
\label{eq:den}
\end{align}
Now, to eliminate the time derivative in the last term of Eq.~(\ref{eq:den}), 
we use the derivative expansion.
Namely, we approximate  
\begin{align}
\lambda T_0\bm{\nabla} \cdot  \partial_0{\vvE}& = \lambda T_0 \bm{\nabla} \cdot h_0^{-1}( -\bm{\nabla} \delP + ... ) \\
                                                &\simeq  -\frac{\lambda T_0}{h_0} \bm{\nabla}^2 \delP ,
\end{align}
where $...$ denotes the second and higher order terms about the derivative, like $\eta \bm{\nabla}^2 \vvE$.
Here, we first used Eq.~(\ref{eq:eckjo}) for $ \partial_0 \vvE$ and next neglect the second-order terms,
which finally yields third-order terms.
Thus, we obtain 
\begin{equation}
\bigl( \partial_0\dele (t, \bm{r}), \deln (t, \bm{r}') \bigr) 
\simeq - h_0\bm{\nabla} \cdot \bigl( \vvE (t, \bm{r}),  \deln (t, \bm{r}')\bigr)
-\lambda \bl (\frac{ n_0 T_0^2}{h_0} \br) \bm{\nabla}^2 \bigl ( \delmu )(t, \bm{r}), \deln (t, \bm{r}')\bigr),
\label{eq:den2}
\end{equation}
where we also used the Gibbs-Duhem relation, Eq.~(\ref{eq:GD}). 
Let us here estimate the equal-time correlations, 
$\bigl( \vvE,  \deln \bigr)$, $\bigl( \vvE,  \dele \bigr)$ and $\bigl( \delmu , \deln \bigr)$.
We assume that the time-reversal property of $\vvE$ $(\vdelj )$ is odd, 
according to Eq.~(\ref{eq:tr2}).
Then, $\vvE$ is orthogonal to $\deln$:
\begin{equation}
\bigl( \vvE , \dele \bigr) = \bigl( \vvE, \deln \bigr) = 0,
\label{eq:orthogonal}
\end{equation}
by the time-reversal symmetry.
On the other hand, $\bigl( \delmu , \deln \bigr)$ turns out to be $\delta ( \bm{r}-\bm{r}')$ from Eq.~(\ref{eq:ort2}).
Finally, we obtain the correlations,
\begin{align}
\bigl( \partial_0\deln(t, \bm{r}),\dele(t, \bm{r}') \bigr) &=0, \\
\bigl( \partial_0\dele(t, \bm{r}) ,\deln(t, \bm{r}') \bigr) &
= -\lambda \bl (\frac{ n_0 T_0^2}{h_0} \br) \bm{\nabla}^2  \delta(\bm{r}-\bm{r}').
\end{align}
We see that the Onsager relation, Eq.~(\ref{eq:onsager}), seems to be violated.
This violation comes from the assumption, Eq.~(\ref{eq:orthogonal}), as will been seen in the following.

Next, let us regard $\vvE $ as a function of $\{ \dele, \delp^i, \deln \}$.
Namely, we consider  $\vdelp $ as an independent variables, instead of $\vdelj $.
Then, we now use Eq.~(\ref{eq:zexpansion}), 
which gives the relation between $ \vdelj$ and $\vdelp $: 
\begin{equation}
\begin{split}
\vdelp &=h_0\vvE-\lambda(T\partial_0\vvL+\bm{\nabla} \delT )\\
&\simeq
h_0\vvE -\lambda\bl (\frac{ n_0 T_0^2}{h_0} \br) \bm{\nabla} \delmu , 
\end{split}
\end{equation}
where we used the derivative expansion and the Gibbs-Duhem relation in the second line.
Solving the above relation about $\vvE$,
 we obtain $\vvE$ as the function of $\{\deln, \dele, \vdelp \}$:
\begin{equation} 
\vvE (\deln, \dele, \vdelp ) = \frac{1}{h_0}\vdelp +\lambda\bl (\frac{ n_0 T_0^2}{h_0^2} \br) \bm{\nabla} \delmu .
\label{eq:LErelation}
\end{equation} 
Here, we note that the time-reversal property of $\vvE ( \dele, \delp^i,\deln ) $
is not odd because those of $\vdelp $ and $\delmu $ are odd and even, respectively.
Substituting Eq.~(\ref{eq:LErelation}) into Eqs.~(\ref{eq:dne}) and (\ref{eq:den2}),
we find
\begin{align}
\bigl( \partial_0\deln (t, \bm{r}),\dele (t, \bm{r}') \bigr) 
& = -\frac{n_0}{h_0} \bm{\nabla} \cdot \bigl( \vdelp (t, \bm{r}),  \dele(t, \bm{r}') \bigr)
-\lambda\bl (\frac{ n_0 T_0}{h_0} \br)^2 \bm{\nabla}^2 \bigl(\delmu (t, \bm{r}), \dele(t, \bm{r}')\bigr), \\
\bigl( \partial_0\dele(t, \bm{r}), \deln(t, \bm{r}') \bigr) 
&=\bigl(\vdelp (t, \bm{r}),\deln (t, \bm{r}')\bigr).
\end{align} 
Here, we note that
$\delmu $ is orthogonal to $\dele$ by Eq.~(\ref{eq:ort1}).
Then, if we assume the time reversal property of $\vdelp $ is odd,
we see that the Onsager relation is satisfied:
\begin{equation}
\bigl( \partial_0\deln(t, \bm{r}),\dele(t, \bm{r}') \bigr) 
= \bigl( \partial_0\dele(t, \bm{r}) ,\deln(t, \bm{r}') \bigr) = 0.
\end{equation}
 
Why cannot we regard $\vdelp$ as a function of $\{ \deln, \dele, \vdelj \}$?
The reason is that $\vdelj $ is not essentially slow,
and then the slow motion of that turn out to be described by the actual slow variables.
Thus, the time reversal property differs from the original operator at the starting point of Sec.~\ref{sec:metric}.
Actually, we can show  that 
the projection of $\vdelj$ on $\{\deln, \dele, \vdelp \}$ gives Eq.~(\ref{eq:LErelation}). 
The derivation is given in the Appendix \ref{sec:projection}.
Namely, the actual expression of Eq.~(\ref{eq:LErelation}) is given as
\begin{equation}
 \cP ( \vdelj (t) )= 
\frac{n_0}{h_0}\vdelp (t) -\lambda\bl (\frac{ n_0 T_0}{h_0} \br)^2 \bm{\nabla }
\bl[ \bl(\frac{\partial(\beta\mu)}{\partial n}\br)_e \deln(t) 
+\bl(\frac{\partial(\beta\mu)}{\partial e}\br)_n \dele(t)\br],
\end{equation}
 where $\cP$ is the projector on $\{ \deln, \dele, \delta \bm{p} \}$,
i.e., $\vdelj $ ($\vvE$) in the Eckart equation is projected one and 
differs from the original operator.  
In consequence, 
the slow variables for the Landau frame actually describe the slow dynamics even for the Eckart frame.

\section{Summary}
\label{sec:summary}
We studied relativistic hydrodynamics by Mori's projection operator method and focused on linear fluctuations around the thermal equilibrium at the rest frame.
From the view point of the projection operator method,
we discussed that the difference of the frames is not the choices of  the reference frames 
but rather those of the slow variables. 
We also found that the the slow variables for the Landau frame are the conserved charges 
whereas those for the Eckart frame include the current of the conserved charge,
which is not essentially slow.
In fact, we derived the slow dynamics by the projection operator method for the sets,
$\{ \dele, \delp^i, \deln \}$ and $\{ \dele, \delp^i, \deln, \delj^i \}$, as the slow variables.
We first showed that the natural choice, Eq.~(\ref{eq:lset}), gives the linearized Landau equations.
Next, we derived the equations of motion for $\{ \dele, \delp^i, \deln, \delj^i \}$,
 which include the Landau and Eckart frames in the view point of the projection operator method.
 And then, we derived the linearized Eckart equation by eliminating $\delp^i$ from the above equations.        
  
We also discussed the slow dynamics is determined only by  $\{ \dele, \delp^i, \deln \}$ and the Landau frame is natural.
In particular, we showed that the equations for $\{ \dele, \delp^i, \deln, \delj^i \}$ has the same modes as for  $\{ \dele, \delp^i, \deln \}$.
Thus, we found that the slow modes are determined only by the Landau's variables.  
Furthermore, by considering the Onsager relation, we illustrated that the slow part of the particle current $\delj^i$ is determined by Landau's variables even for the Eckart equation.
Recently, some authors also point out 
that the Landau frame is natural for the relativistic hydrodynamics,
based on the renormalization group method \cite{Tsumura:2012ss}.

Here, we note that this study is first for  
the derivation of relativistic hydrodynamics based on the projection operator method.  
We stress that our derivation is independent of the microscopic details; 
we determine the metric 
from the Lorentz symmetry on microscopic scale and the thermodynamics.
The earlier studies \cite{Denicol,Van,Muronga,Tsumura:2011cj} 
assume the relativistic Boltzmann equation as the underlying microscopic theory,
which is, however, only valid for a weakly-correlated system like a dilute gas. 
Furthermore, we note that our study is independent of what are the local equilibrium and the local rest in the relativistic fluids
because we considered the fluctuations from globally equilibrium state.  Instead, our study is restricted in the linear fluctuations from the equilibrium state at rest.   
Now, we comment on the Lorentz covariance of  linearized hydrodynamics.
The linearized Landau equations, (\ref{eq: ln})-(\ref{eq: lj}), are not Lorentz covariant.
The reason is the following:
We here considered the fluctuations in the background medium.
For such a system, the Lorentz transformation boosts the fluctuations but does not the background.
Then, the boosted system differs from that before the boost.
Thus, the linearized equations are valid only for the rest frame of the medium and not Lorentz covariant.
Actually, by the same reason, the Navier-Stokes equation is Galilei covariant whereas 
the linearized one is not covariant.

In this paper, we used the linear projection operator, and 
our study is restricted in the linear regime.
Then, it is interesting to derive relativistic hydrodynamics
 by the nonlinear projection operator \cite{mori:fujisaka}.
We note that, for nonrelativistic fluids, the full Navier-Stokes equation is derived 
by the nonlinear one \cite{kim}.
Furthermore, we here focused and discussed the Landau and Eckart frames.
Then, we concluded that the Landau frame is natural  for the slow dynamics.
 However, other problems of relativistic hydrodynamics such as the acausal propagation
 although it is studied from the macroscopic point of view \cite{Denicol:2008ha},
 have not been well understood from the underlying microscopic theory yet. 
It would be interesting to discuss these in the view point of the projection operator method.
The projection operator method may give an insight into these problems, as well as that of the frames.

\acknowledgements
Y.H. thank Teiji Kunihiro for useful discussions.
This work was supported by JSPS KAKENHI Grant Numbers 23340067 and 24740184.

\appendix

\section{Properties of inner product} \label{sec:innerProduct}
Here, we summarize properties of the inner product defined in Eq.~(\ref{eq:innerProduct}).
For Hermitian operators,
\begin{align}
  (\hA(t), \hB(0))&=(\hA(t), \hB (0))^*= (\hB (0),\hA(t)), \label{eq:innerProductReal}\\
(i\Lv \hA(t),\hB (0)) &= -(\hA(t), i\Lv \hB (0)), \label{eq:innerProductByPart}\\
 (i\Lv \hA(t), \hB (0))&=-\frac{i}{\beta}\average{ [\hA(t),\hB (0)] }, \label{eq:KuboIdentity}\\
(\hA(t), \hB (0)) &= \epsilon_{A}\epsilon_{B}(\hA(-t), \hB (0)) = \epsilon_{A}\epsilon_{B}(\hB (t),\hA (0)) \label{eq:innerProductTimeReversal}
\end{align}
are satisfied.
Here $\epsilon_{A}$ and $\epsilon_{B}$ denote the sign associated with time reversal transformation, which is defined with the time reversal operator $\mathcal{T} $ as
$\mathcal{T}^{-1} \hA(t) \mathcal{T} = \epsilon_{A} \hA(-t)$.

\section{Memory function}\label{sec:MemoryFunction}
In this Appendix, we derive the full expression of the memory function following Ref.~\cite{Koide:2008nw}.
Let us first define the key functions:
\begin{align}
\dot{\varXi}(t)&\equiv(i\Lv \hA_n(t), \hA^m),\\
\ddot{\varXi}(t)&\equiv((i\Lv)^2 \hA_n(t), \hA^m).
\end{align}
The reversible term can be expressed by $i\varOmega=\dot{\varXi}(0)$.
$\ddot{\varXi}(t)$ coincides with the memory function obtained by replacing $\cQ$ with unity.
Expanding $\exp({t \cQ i\Lv})$ in terms of $\cP i\Lv$,
we obtain
\begin{equation}
\begin{split}
e^{t\cQ i\Lv}
&=e^{it\Lv}\Bigl(1+ \sum_{n=1}^\infty(-1)^n\int_0^t dt_1 \cdots \int _0^{t_{n-1}}dt_n
i\Lv^ P(t_1)\cdots i\Lv^ P(t_{n}) 
\Bigr),
\end{split}
\end{equation}
where 
\begin{equation}
\Lv^P(t)\equiv e^{-i\Lv t}\cP \Lv e^{i\Lv t}.
\end{equation}
Acting  $\Lv^P(t)$ to $\hA_n$, we obtain
\begin{equation}
i\Lv^P(t)\hA_{m}(-t')= e^{-i\Lv t}Pi\Lv e^{i\Lv t}\hA_{m}(-t')=
\hA_{n}(-t)(i\Lv \hA_m(t-t'),\hA^{n})= {\dot{\varXi}_m}{}^n(t-t')\hA_{n}(-t).
\end{equation}
Then, for an operator $\hO$, 
\begin{equation}
\begin{split}
e^{t\cQ i\Lv} \hO
&=e^{it\Lv}\Bigl(1+ \sum_{n=1}^\infty(-1)^n\int_0^t dt_1 \cdots \int _0^{t_{n-1}}dt_n
i\Lv^P(t_1)\cdots i\Lv^P(t_{n})
\Bigr)\hO\\
&=\hO (t)+ \sum_{n=1}^\infty(-1)^n\int_0^t dt_1 \cdots \int _0^{t_{n-1}}dt_n\\
&\qquad\times\dot{\hO }_P(t_n)\dot{\varXi}(t_{n-1}-t_n)\dot{\varXi}(t_{n-2}-t_{n-1})\cdots \dot{\varXi}(t_1-t_{2})
\hA(t-t_1)
\end{split}
\end{equation}
is satisfied. Here we defined
\begin{equation}
\dot{\hO }_P(t)=(i\Lv\hO (t),\hA^m).
\end{equation}
Performing the Laplace transform,
we obtain
\begin{equation}
\begin{split}
\int dt e^{-tz}e^{t\cQ i\Lv} \hO
&=\hO (z)-\dot{\hO}_P(z) \frac{1}{1+\dot{\varXi}(z)} \hA(z).
\end{split}
\end{equation}
Using this equation, we find the full expressions of  the nose term and the memory function in the Fourier-Laplace space:
\begin{align}
\noise(z,\bm{k})&=i\Lv \hA(z,\bm{k})-(i\varOmega+\ddot{\varXi}(z,\bm{k}))\frac{1}{1+\dot{\varXi}(z,\bm{k})}\hA(z,\bm{k}),\label{eq:noiseFunction}\\
\memory(z,\bm{k})&= - (\ddot{\varXi}(z,\bm{k})-i\varOmega(\bm{k})\dot{\varXi}(z,\bm{k}))\frac{1}{1+\dot{\varXi}(z,\bm{k})}.
\label{eq:memoryFunction}
\end{align}

\section{Projection of $\delta \bm{j}$ on $ \{\dele, \delta \bm{p}, \deln \}$ }
\label{sec:projection}
 Here, we show that the projection of $\vdelj$ on
$\{ \deln,\dele, \vdelp \}$ gives Eq.~(\ref{eq:LErelation}).
Namely, we will show 
\begin{equation}
 \cP \bl[ \delj (t,\bm{x}) 
-\frac{n_0}{h_0}\vdelp (t,\bm{x}) +\lambda\bl (\frac{ n_0 T_0}{h_0} \br)^2 \bm{\nabla} \delmu (t,\bm{x})] \br] =0,
\label{eq:ProjectionJ}
\end{equation}
to the first order in $k$.
In the Fourier-Laplace space, Eq.~(\ref{eq:ProjectionJ}) becomes
\begin{equation}
 \cP \bl[ \vdelj(z,\bm{k}) 
-\frac{n_0}{h_0}\vdelp (z,\bm{k}) +i\bm{k}\lambda\bl (\frac{ n_0 T_0}{h_0} \br)^2  \delmu (z,\bm{k})] \br] =0.
\label{eq:migeruFLT}
\end{equation}
Here we decompose Eq.~(\ref{eq:migeruFLT}) into the longitudinal and transverse components:
\begin{align}
\cP \bl[ \delj_{\parallel}(z,\bm{k}) -\frac{n_0}{h_0}\delp_{\parallel} (z,\bm{k})
 +ik\lambda\bl (\frac{ n_0 T_0}{h_0} \br)^2  \delmu (z,\bm{k})] \br] &=0, 
\label{eq:migeruL}\\
\cP \bl[ \vdelj_\perp (z,\bm{k}) -  \frac{n_0}{h_0}\vdelp_\perp (z,\bm{k}) \br] & =0.
\label{eq:migeruT}
\end{align}

Let us show the equation for the transverse component, Eq.~(\ref{eq:migeruT}).
Consider the projections, $ \cP \vdelj_\perp (z,\bm{k})$ and $\cP \vdelp_\perp (z,\bm{k})$,
which are given as 
\begin{align}
\cP \vdelj_{\perp} (z,\bm{k}) &= { \varXi_{j_{\perp}} }^{n}(z,\bm{k})\deln (0,\bm{k})+{\varXi_{j_{\perp}}}^{e}(z,\bm{k}) \dele (0,\bm{k})+  {\varXi_{j_{\perp}}}^{p_{\perp}}(z,\bm{k}) \vdelp_{\perp}(0,\bm{k}), \\
\cP \vdelp_{\perp} (z,\bm{k}) &= {\varXi_{p_{\perp}}}^{n}(z,\bm{k}) \deln (0,\bm{k})+{\varXi_{p_{\perp}}}^{e}(z,\bm{k}) \dele (0,\bm{k})+  {\varXi_{p_{\perp}}}^{p_{\perp}}(z,\bm{k}) \vdelp_{\perp}(0,\bm{k}).               
\end{align}
We note that $\bm{k}$ expansions of ${\varXi_{j_{\perp}}}^{n}(z,\bm{k})$
and ${\varXi_{j_{\perp}}}^{e}(z,\bm{k})$ give only odd-order terms in $\bm{k}$ from
the tensor structure.
Then,  we can drop ${\varXi_{j_{\perp}}}^{n}(z,\bm{k})$ and ${\varXi_{j_{\perp}}}^{e}(z,\bm{k})$ 
because the odd terms are orthogonal to the transverse component. 
Then, Eq~(\ref{eq:migeruT}) turns out to be 
\begin{equation}
\bl[ {\varXi_{j_{\perp}}}^{p_{\perp}}(z,\bm{k}) -\frac{n_0}{h_0}{\varXi_{p_{\perp}}}^{p_{\perp}}(z, \bm{k}) \br] \vdelp_{\perp}(0,\bm{k}) =0.
\end{equation}

Here, let us consider $ {\varXi_{j_{\perp}}}^{p_{\perp}}(z,\bm{k}) $ and 
$ {\varXi_{p_{\perp}}}^{p_{\perp}}(z,\bm{k}) $.
For this task, we now use the equations of motion for $\{ \deln, \dele, \vdelp, \vdelj \}$, Eqs.~(\ref{eq:EEPOM}), (\ref{eq:ENPOM}), (\ref{eq:eomP}) and (\ref{eq:eomJ}). 
From these equations,  we obtain the equations for the transverse components as 
\begin{equation}
\begin{pmatrix}
z+k^2\varGamma_{\perp } & 0\\
z+\varGamma_j & -(h_0/n_0)\varGamma_j
\end{pmatrix}
\begin{pmatrix}
&\vdelp_{\perp  } (z, \bm{k}) \\ 
&\vdelj_{\perp  } (z, \bm{k}) \\
\end{pmatrix}
=
\begin{pmatrix}
\vdelp_{\perp }(t=0,\bm{k}) \\
 \vdelp_{\perp }(t=0,\bm{k})
+n_0T_0g^{jj}\bl(\vdelj_{\perp }(t=0,\bm{k})-\frac{n_0}{h_0}\vdelp_{\perp }(t=0,\bm{k})\br)
\end{pmatrix}.
\end{equation}
Therefore,
\begin{align}
\vdelp_\perp(z,\bm{k})&=\frac{1}{z+k^2\varGamma_{\perp }}\vdelp_\perp (t=0,\bm{k}), \\
\vdelj_\perp(z,\bm{k})&=\frac{(n_0/h_0)}{z+k^2\varGamma_{\perp }}\vdelp_\perp (t=0,\bm{k})-\frac{ n_0^2T_0g^{jj}}{h_0\varGamma_j}
\bl(\vdelj_{\perp }(t=0,\bm{k})-\frac{n_0}{h_0}\vdelp_{\perp }(t=0,\bm{k})\br) .
\end{align}
If we notice that the conjugate variable for $\delp^i $ is  
\begin{equation}
\hA^{p^i} = \frac{1}{h_0T_0}\delp^i,
\end{equation}
in the space of $\{ \dele, \vdelp, \deln \}$,
we find 
\begin{align}
{\varXi_{j_{\perp}}}^{p_{\perp}}(z,\bm{k}) &= \frac{(n_0/h_0)}{z+k^2\varGamma_{\perp }},\\
{\varXi_{p_{\perp}}}^{p_{\perp}}(z,\bm{k}) &= \frac{1}{z+k^2\varGamma_{\perp }},  
\end{align}
where we used Eqs.~(\ref{eq:gpp}) and (\ref{eq:gpj}). 
Finally, we arrive at 
\begin{equation}
\bl[ {\varXi_{j_{\perp}}}^{p_{\perp}}(z,\bm{k}) 
-\frac{n_0}{h_0}{\varXi_{p_{\perp}}}^{p_{\perp}}(z, \bm{k}) \br]  =0.
\end{equation}
We have shown the equation for the longitudinal component, Eq.~(\ref{eq:migeruT}).
By the similar procedure, we can show that for the longitudinal component, Eq.~(\ref{eq:migeruL}).

\end{document}